\documentclass[[11pt]{article}
\usepackage[dvips]{graphicx}  
\begin{document}

\title{\LARGE \bf   Self-Referential Noise as a 
Fundamental Aspect of Reality
\thanks{Contribution to the 2nd International Conference on {\bf Unsolved Problems of Noise},
Adelaide 1999.}}  
\author{{Reginald T. Cahill  and     Christopher M. Klinger}\\
  {Department of Physics, Flinders University
\thanks{E-mail: Reg.Cahill@flinders.edu.au, Chris.Klinger@flinders.edu.au}}\\ { GPO Box
2100, Adelaide 5001, Australia }}

\date{}
\maketitle

\begin{abstract}
Noise is often used in the study of open systems, such as in classical Brownian motion  and
in  Quantum Dynamics, to model the influence of the environment. 
However generalising results from G\"{o}del and  Chaitin in mathematics suggests that  systems that
are sufficiently rich that self-referencing is possible contain intrinsic randomness.  We argue that
this is relevant to modelling the universe, even though it is by definition a closed system. We show
how a three-dimensional process-space may arise, as a Prigogine dissipative structure, from a non-geometric
order-disorder model driven by, what is termed, self-referential noise.

\end{abstract}

\section*{\bf  Introduction\label{section:Introduction}}
For over 300 years theoretical physics has very successfully modelled reality using geometrical models of
the phenomena of space and time, and with deterministic fields and objects attached to the geometrical
object which we call space. 
 However there are indications from
the quantum theory that there are processes, as  in the Aspect experimental
study\cite{Aspect} of the Einstein-Podolsky-Rosenfeld(EPR) effect (see 
Bell\cite{Bell} for discussion), that   a fundamental non-local
random connectedness is needed to understand  the quantum measurement
process. 
For this and other reasons there have been attempts to construct more
fundamental models of reality that do not begin with the assumption of an
{\it a priori} geometry, and which are known as pregeometric
models\cite{Wheeler80}.  However, even extant pregeometric modellings of
reality have had no success in explaining  the phenomenon of {\it
space}, and in particular why {\it space} is effectively {\it three
dimensional} for most phenomena.  As well the strong
belief in the universality of the deterministic time evolution of physical
systems  (with the exception perhaps of the quantum measurement processes)
does not take account of fundamental limitations that first began to
appear with the discoveries of G\"{o}del in mathematical logic. 
Subsequent developments of G\"{o}del's theorems by Chaitin\cite{Chaitin}
led to the discovery that mathematical systems sufficiently rich that
self-referencing is possible contain intrinsic randomness. This appears to
indicate a fundamental dichotomy between the limitations indicated by
mathematic logic and the assumption  of absolute determinism in
theoretical physics.  We argue that the resolution of this dichotomy  is
relevant to modelling the universe, and we   show how a three-dimensional
process-space may arise from a non-geometric order-disorder model as a
Prigogine\cite{Prigogine} dissipative structure driven by, what is
termed,   self-referential noise (SRN).  We call this noise SRN  to indicate its
relationship to G\"{o}del's  theorems. However, note that this noise is not
itself self-referential and nor is the model considered herein. Hence while noise is often used in
the study of open systems, such as in classical Brownian motion  and in 
Quantum Dynamics\cite{IP}, to model the influence of the environment, here
we argue that in the case of the universe, which by definition is a closed
system, we must nevetheless use noise  - not to take account of an {\it
environment}, but to model the limitations indicated by
logic\cite{CK97,CK98}.  Patton and  Wheeler\cite{PW}
conjectured some time ago that G\"{o}del's results in mathematics might be
relevant to understanding cosmogony.  For
reasons discussed elsewhere\cite{CK97,CK98} we call this system a
Heraclitean Process System (HPS).
 
\section*{Process Systems\label{section:Heraclitean-Process-System}}  
Modelling reality  at a fundamental level faces the problem of what to begin with?   In \cite{CK97,CK98}
we proposed a resolution to this problem by appealing to the phenomenon of self-organising criticality
(SOC)\cite{SOC}.  In the proposed bootstrap model  start-up components (called monads)
acquire a self-consistent meaning only as we reveal the fractal structures (i.e. criticality) that emerge. SOC
systems have the property of {\it universality}, i.e. the behaviour of the system at  a self-organised 
critical point is not uniquely characteristic  of individual systems.
Smolin\cite{Smolin} has discussed the possible relevance of
SOC  to cosmology.

The construction of a viable HPS can only be achieved at present by inspired guessing based in
part upon the lessons of Quantum Field Theory (QFT)(see \cite{CK96})  and   
  the peculiarities of the quantum measurement process which indicate
 manifestations of  non-linear and non-local random processes\cite{IP}. 
Our first HPS-SOC model is described by a non-linear noisy iterative map\cite{CK97,CK98}, where the
parameters $\alpha, \beta$ and $\gamma$  and the matrix $B_c$, see (2a), simplify the analysis:   
\begin{equation}
B_{ij} \rightarrow B_{ij} -\alpha(\beta^{-2} B+B^{-1}+\gamma B_c)_{ij} + w_{ij},  \mbox{\ \ } i,j=1,2,...,2M;
M
\rightarrow
\infty,
\label{eq:map}\end{equation}

We introduce, for convenience only, some terminology: we think of
$B_{ij}$ as indicating the connectivity or relational strength between two monads
$i$ and $j$ (these monads acquire a meaning
later). The monads concept was introduced by Leibniz, who espoused the {\it relational}
mode of thinking in response to and in contrast with Newton's {\it
absolute} space. It is important to note that the 
 iterations of the map do not constitute {\it a priori} the phenomenon of time, since they are
to perform the function of producing the needed fractal structure which characterises universality in
SOC. It is significant that this HPS is non-quantum, with the quantum    phenomenon  being  emergent,
see\cite{CK97}.
\begin{equation}B_c=\left(\begin{array}{rrrrrr}
0 & +1 & 0 & 0\\
-1 & 0 & 0 & 0 \\
0 & 0 & 0 & +1 \\
0 & 0 & -1 & 0 & \\
 &&&&. \\ &&&&&. \\          
\end{array}\right)(a),\mbox{\ \ \ } B=\left(\begin{array}{rrrrrr}
g_1 &  & & \\
 & \mbox{\Huge{\Huge{$\bigcirc$}}} &  &  \\
 &  & g_2 &  \\
 &  &  & g_3 \\
       &&&&c_1 \\&&&&&c_2 \\           
\end{array}\right)(b).\label{eq:arrays}\end{equation}

 The monad  $i$ has a pattern of  dominant (larger valued $B_{ij}$) connections
$B_{i1}, B_{i2},...$, where
$B_{ij}=-B_{ji}$  avoids  self-connection ($B_{ii}=0$), and real number valued. The self-referential noise
 $w_{ij}=-w_{ji}$ are
independent random variables  for each $ij$ and for each iteration, and  with variance $\eta$. 
Parameters  satisfying  $\alpha > \beta  \gg \eta  \gg \gamma$ result in  identifiable emergent
and evolving  patterns.
   With the noise absent the iterator converges to
 the  {\it condensate} $\beta B_c$ (but with $\gamma=0$ to one of $\beta RB_cR^{-1}$ where the
matrix $R$ depends on the initial $B$).  This behaviour is similar to the condensate of Cooper
pairs  in QFT\cite{CK96}, but here the condensate (indicating a non-zero dominant configuration) does not
have any space-like structure. However in the presence of the noise,   after an
initial chaotic behaviour when starting the iterator from
$B\approx 0$, the dominant mode is the formation of a randomised condensate  $C\approx
\mu\otimes B_c + B_b$,  indicating  $B_c$ but with the $\pm 1's$ replaced by $\pm\mu_i$'s   (where  the
$\mu_i$ are small and given by a computable iteration-dependent probability distribution ${\cal M}(\mu)$)  and
with a noisy background
$B_b$ of very small   $B_{ij}$. 

The key discovery is that there is an extremely small (relative to $M$) self-organising process
buried within this  condensate  and which has the form of a three-dimensional  fractal process-space,
which we now briefly explain. This structure  is an example of a Prigogine far-from-equilibrium dissipative
structure\cite{Prigogine}, emerging from the unstructured condensate and driven by the SRN.
 Under the mapping the noise term
will produce rare large value
$B_{ij}$, and  these  $B_{ij}$ will persist under the mapping  (through more
iterations than smaller valued $B_{ij}$) and form fluctuating patterns of connections, whose structure  
 we now identify.  
\newline\indent Consider the connectivity  from the point of
view of one monad, call it monad $i$.  Monad $i$ is connected via these large $B_{ij}$ to a
number of other monads, and the whole set of connected monads forms a tree-graph relationship. This is
because  the large links are very improbable, and a tree-graph relationship is much more probable
than a similar graph involving the same monads but with additional links. The set of all large
valued $B_{ij}$ then form tree-graphs disconnected from one-another; see Fig.1a.  In any one
tree-graph the natural `distance' measure for any two monads within a graph is  the smallest number
of links  connecting  them. 
Let
$D_1, D_2,...,D_L$ be the number of nodes of distance
$1,2,....,L$ from monad $i$ (define $D_0=1$ for convenience), where $L$ is the largest distance
from $i$ 
in a particular tree-graph, and let $N$ be the total number of nodes in the tree. 
See Fig.1b for an example.

 Now consider
the number ${\cal N}(D,N)$ of different $N$-node trees, with the same distance distribution
$\{D_k\}$, to which $i$ can belong.  By counting\cite{CK98} the different linkage patterns, together with
permutations of the monads we obtain (3). We may compute the most likely tree-graph
structure by  maximising   ${\cal N}(D,N)$ with respect to $\{D_k\}$.   Fig.2  shows a typical  result. 
\begin{equation}
{\cal N}(D,N)=\frac{(2M-1)!D_1^{D_2}D_2^{D_3}...D_{L-1}^{D_L}}{(2M-N)!D_1!D_2!...D_L!},
\label{eq:analytic}\end{equation}
 Also shown is the approximate analytic form\cite{Nagels}   $D_k=\frac{2N}{L}\sin^2(\pi k/L)$.  
These results imply  that the most
likely tree-graph structure to which a monad can belong  has a distance
distribution $\{D_k\}$  which indicates that the tree-graph is embeddable in
a 3-dimensional hypersphere, $S^3$. 

\vspace{3mm}
\hspace{-18mm}\begin{minipage}[t]{60mm}
\hspace{15mm}\includegraphics[scale=0.4]{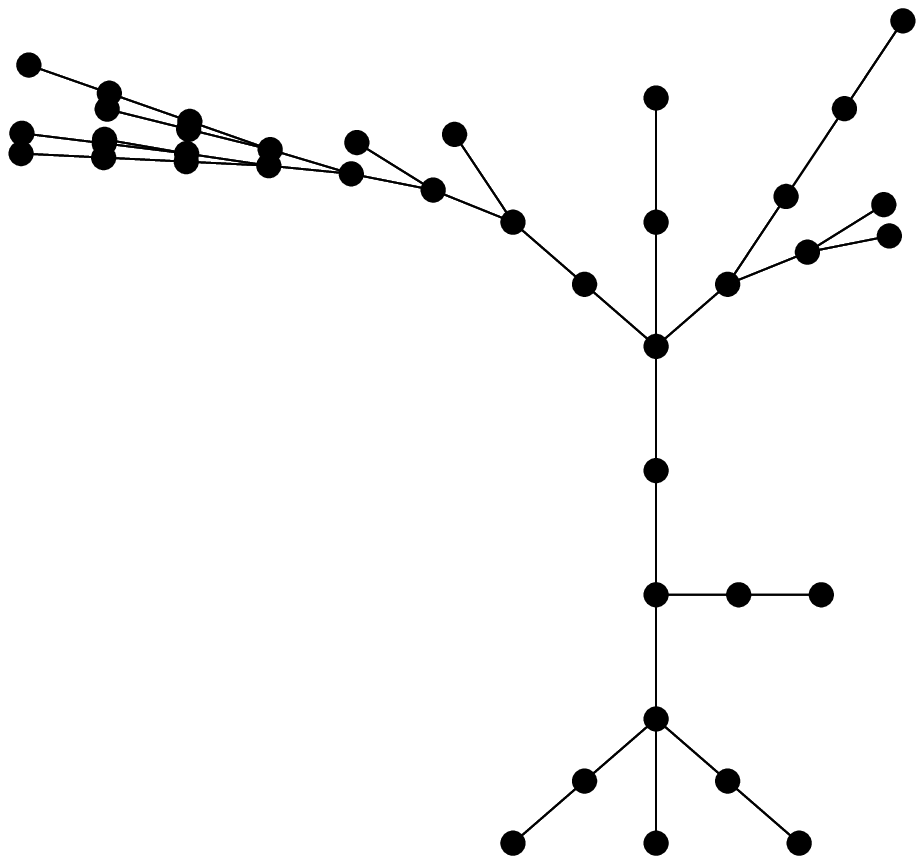}
\makebox[130mm][c]{(a)}
\end{minipage}
\begin{minipage}[t]{60mm}
\hspace{10mm}\includegraphics[scale=0.4]{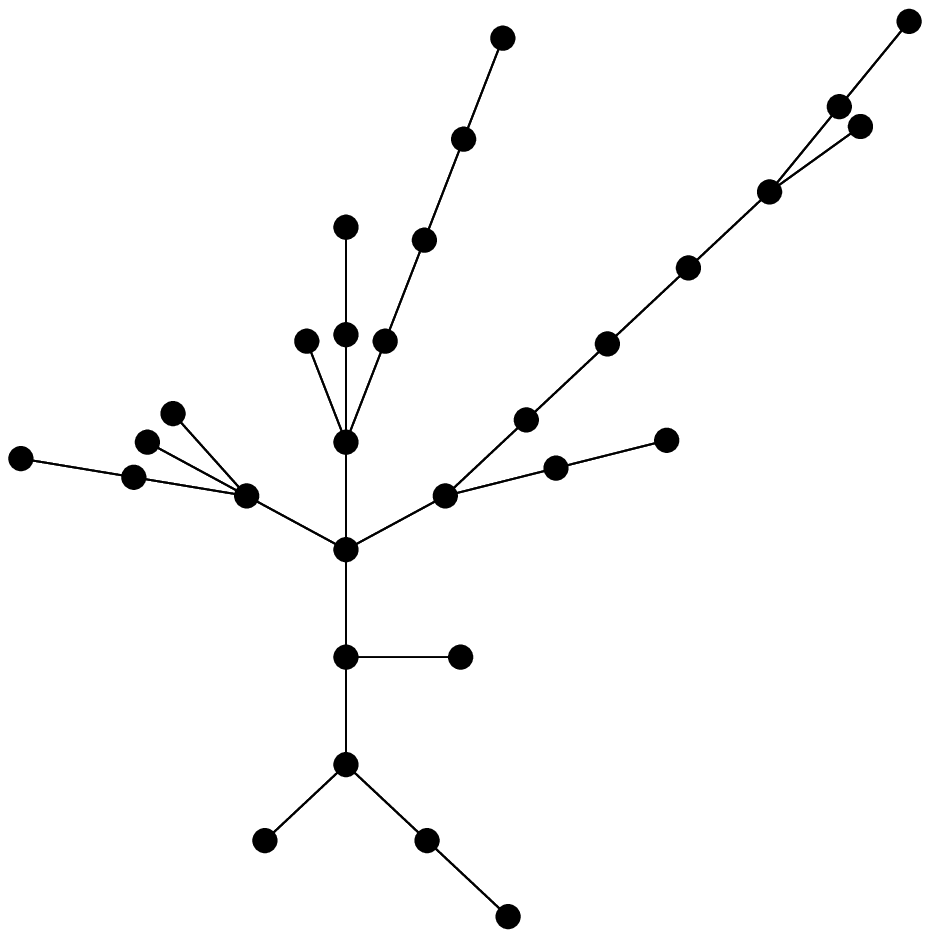}
\end{minipage}
\begin{minipage}[t]{50mm}
\setlength{\unitlength}{0.20mm}

\hspace{3mm}\begin{picture}(0,200)(100,0)  
\thicklines

\put(155,205){\line(3,-5){60}}
\put(155,205){\line(-3,-5){60}}
\put(115,140){\line(3,-5){42}}
\put(195,140){\line(-3,-5){21}}

\put(135,200){ \bf $i$}
\put(225,200){ \bf $D_0\equiv 1$}
\put(225,140){ \bf $D_1=2$}
\put(225,100){ \bf $D_2=4$}
\put(225,65){ \bf $D_3=1$}

\put(155,205){\circle*{5}}

\put(115,140){\circle*{5}}
\put(195,140){\circle*{5}}

\put(95,105){\circle*{5}}
\put(135,105){\circle*{5}}
\put(175,105){\circle*{5}}
\put(215,105){\circle*{5}}

\put(155,70){\circle*{5}}
\end{picture}

\makebox[25mm][c]{(b)}
\end{minipage}
\begin{figure}[ht]
\vspace{-5mm}\caption{\small (a) Rare and large components of $B$ form disconnected
tree-graphs, (b) An $N=8$ tree-graph with $L=3$ for monad {\it i}, with indicated distance
distribution $D_k$.
 \label{figure:gebits}}
\end{figure}

\vspace{-5mm}
\begin{figure}[ht] 
\hspace{45mm}\includegraphics[scale=0.751]{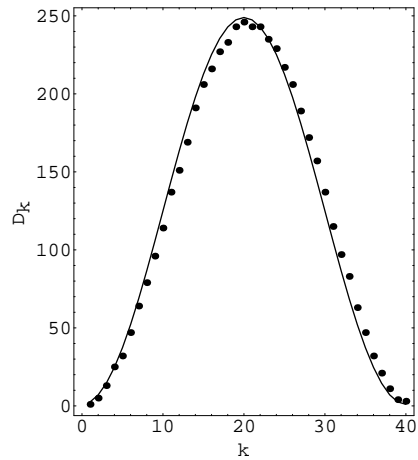}
\caption{\small  Points show the $D_k$ set and $L=40$ value found by numerically  maximising
(\ref{eq:analytic}) for fixed  $N=5000$. Curve shows
$D(k)=\frac{2N}{L}\sin^{d-1}(\frac{\pi k}{L})$ with $d=3$ and $L=40$, showing  excellent
agreement, and indicating a weak embeddability in $S^3$.
 \label{figure:Plot}}
\end{figure}

\vspace{0mm}

   We call these
tree-graph $B$-sets {\it gebits} (geometrical bits). However $S^3$  
embeddability of these gebits is a weaker result than demonstrating the necessary emergence of  $S^3$-spaces,
since extra cross-linking connections would be required   for this to produce a strong embeddability; for evidence
of this see \cite{CK98}.

The monads for which the $B_{ij}$ are, from the SRN term,  large thus form  disconnected
 gebits, and in (2b) we relabel the monads to bring these new gebits $g_1,g_2,g_3,..$ to
block diagonal form, with the remainder indicating the small and growing thermalised
condensate, $C=c_1\oplus c_2\oplus c_3\oplus...$   In 2b the $g_i$
indicate unconnected gebits, while the icon $\bigcirc$ represents older and connected gebits,
and suggests  a compact 3-space (see below). The remaining very small
$B_{mn}$, not shown in (2b), are background  noise only.

A key dynamical feature is that  most gebit  matrices $g$ have $\mbox{det}(g) =0$, since most tree-graph
connectivity matrices are degenerate\cite{CK98}.   These
$\mbox{det}(g) =0$ gebits  form a   {\it reactive gebits}  subclass (i.e.  in the presence
of background noise $(g_1\oplus g_2\oplus g_3\oplus..)^{-1}$ is well-defined and some
elements  large)   of all those gebits generated by the SRN, and they are the building
blocks of the dissipative structure.  The self-assembly process is as follows: before the
formation  of the thermalised condensate
$B^{-1}$ generates new connections (large $B_{ij}$) almost exclusively between  gebits
and the remaining non-gebit sub-block (having det$\approx 0$ but because here all
the involved
$B_{ij}\approx 0$), resulting in the decay, without gebit interconnection, of each
gebit.  However once the condensate has formed (essentially once the system has `cooled'
sufficiently) the condensate
$C=c_1\oplus c_2\oplus c_3\oplus...$ acts as a quasi-stable (i.e.
$\mbox{det}(C)=\prod_i
\mbox{det}(c_i)\neq0$) sub-block of (2b) and the  sub-block of gebits may be inverted separately. The  gebits are
then interconnected (with many gebits present cross-links are more probable than self-links)
via new links formed by
$B^{-1}$, resulting in the larger structure indicated by the $\bigcirc$ in (2b). Essentially, in the presence of
the  condensate, the gebits are {\it sticky}. 

 Continuing
studies\cite{CK97,CK98} suggest that this network of self-assembling gebits forms a three-dimensional fractal
process-space (the $\bigcirc$ in (2b) - essentially a Prigogine dissipative structure): fractal because
sub-networks of gebits are themselves formed into larger networks.  It is this rapidly expanding process-space that
we associate with the phenomenon of {\it space}, and from the  endophysics of this space  the condensate is
completely non-local. It is also clear, finally,  that the original monads can be interpreted as themselves being
networks of connected gebits. For this reason we thus have  a bootstrap HPS\cite{CK98} (i.e. the start-up
components are identical in form to emergent components).   After a   transient regime of expansion,
dominated by the interaction of  topological defects produced during the early formation phase of the 
process-space, one would expect the process-space to undergo an exponential expansion because the growth in
the number $n$ of  gebits  within the process-space would be  described by a growth-decay equation
\begin{equation}
\frac{d n}{dt}= an-bn.
\label{eq:exp}\end{equation} 
This suggest that the HPS model may provide an explanation for the cosmological constant 
 which now appears to be firmly established from observational evidence\cite{cosmo1,cosmo2}.  

A process-space is   not equivalent to an inert geometrical
space.  In particular this implies a finite speed of propagation of any disturbance through the process-space and
other distortion effects, caused by the need for the disturbance to be processed by the formation, interconnection
and finally decay of the gebits. Toffoli\cite{Toffoli} has speculated about such phenomena and its possible
explanation of General Relativity-like effects within the area of Cellular Automata.   

\section*{Conclusions and Open Questions\label{section:Conclusion}} 
We have briefly discussed  the  problem which arises when we attempt to model and comprehend
the universe as a closed system without assuming high level phenomena such as space and  time.
     Our analysis is based upon the  notion that a closed
self-referential system, and  the universe is {\it ipso facto} our only true instance,  is necessarily noisy.
 This follows as a conjectured generalisation of the work of G\"{o}del and Chaitin on self-referencing in
the abstract and artificial game  of mathematics.  To explore the implications we have considered a {\it
non-quantum non-geometric  non-linear noisy iterative map}.  The analysis of this map shows that the first
self-organised structure to arise  is a dynamical Prigogine-like dissipative  process-3-space formed from
interconnecting pieces of 3-geometry - the gebits.    We suggest that the concept of a non-local intrinsic  noise
has been a  major missing component of traditional modelling of reality.  As discussed
elsewhere\cite{CK97,CK98} this model also generates the phenomenon of the {\it present
moment effect} - an effect missing from the Newtonian and Einsteinian  geometrical
models, and an {\it objectification process} related to the phenomenon of ({\it
classical}) objects and  to the behaviour of quantum detectors. 

Both analytical and numerical studies have indicated that the interconnecting gebits  form a complex
dynamical network once the system has cooled sufficiently for the non-local condensate to have formed. However a
key open question is the proof, probably analytical,  that this network is indeed three-dimensional.  If this
conjecture is confirmed then we would have for the first time a model which predicts the emergence of 
the complex phenomenon of {\it space}, and one that is richer than that  used in the present day geometric
modelling  in physics.

We thank Dr Derek Abbott and members of the UPoN99 Conference Secretariat for their efforts in organising
this conference.

\end{document}